\documentclass[prb,twocolumn,superscriptaddress,showpacs,amsmath,amssymb]{revtex4}
\usepackage{amsfonts}
\usepackage{bm}
\usepackage{verbatim}

\usepackage{graphicx}

\def\Z{\mathbb{Z}}

\begin{document}
\title{Andreev-Majorana bound states in superfluids.}

\author{Mikhail~Silaev}
\affiliation{Low Temperature Laboratory, Aalto University,  P.O. Box 15100, FI-00076 Aalto, Finland}
\affiliation{Institute for Physics of Microstructures RAS, 603950 Nizhny Novgorod, Russia}

\author{G.E.~Volovik}
\affiliation{Low Temperature Laboratory, Aalto University,  P.O. Box 15100, FI-00076 Aalto, Finland}
\affiliation{Landau Institute for Theoretical Physics, acad. Semyonov av., 1a, 142432,
Chernogolovka, Russia}

\date{\today}

\begin{abstract}
We consider Andreev-Majorana (AM) bound states with zero energy on surfaces, interfaces and vortices in
different phases of the $p$-wave superfluids. We discuss the chiral superfluid $^3$He-A, and time reversal invariant phases: superfluid $^3$He-B, planar and polar phases.
The AM zero modes are determined by topology in bulk, and they disappear at the quantum phase transition from the topological
to non-topological state of the superfluid. The topology demonstrates the interplay of dimensions.
In particular, the zero-dimensional Weyl points in chiral superfluids (the Berry phase monopoles in momentum space) give rise to the one-dimensional  Fermi arc of AM bound states on the surface and to the one-dimensional 
flat band of AM modes in the vortex core. The one-dimensional  nodal line in the polar phase produces the  two-dimensional flat band of AM modes on the surface. 
The interplay of dimensions also connects the AM states in superfluids with different dimensions.
For example, the topological properties of the spectrum of bound states in the three-dimensional $^3$He-B is connected to the properties of the spectrum in
the two-dimensional planar phase (thin film).

\end{abstract}
\maketitle

\tableofcontents

 \section{Introduction}

Majorana fermions are ubiquitous for superconductors and fermionic
superfluids. The Bogoliubov- de Gennes
 equation for fermionic Bogoliubov-Nambu quasiparticles can be brought to a real form
by unitary transformation. This implies the linear
relation  between the particle and antiparticle
field operators, which is the hallmark of a Majorana
fermion. The fermionic statistics and Cooper pair correlations  give rise to
Majorana fermions, irrespective of  geometry, dimensionality, symmetry and topology \cite{Beenakker2014,Chamon2010,Senthil2000}.
The role of topology is to protect gapless Majorana fermions, which play the major role
at low temperature, when the gapped degrees of freedom are frozen out.
For some combinations of geometry, dimensionality and symmetry these Majorana fermions behave as
emergent massless relativistic particles. This suggests that Majorana fermions may serve as
building blocks for construction of the Weyl particles of Standard Model  \cite{VolovikZubkov2014}.

\begin{figure}[hbt]
\centerline{\includegraphics[width=0.90\linewidth]{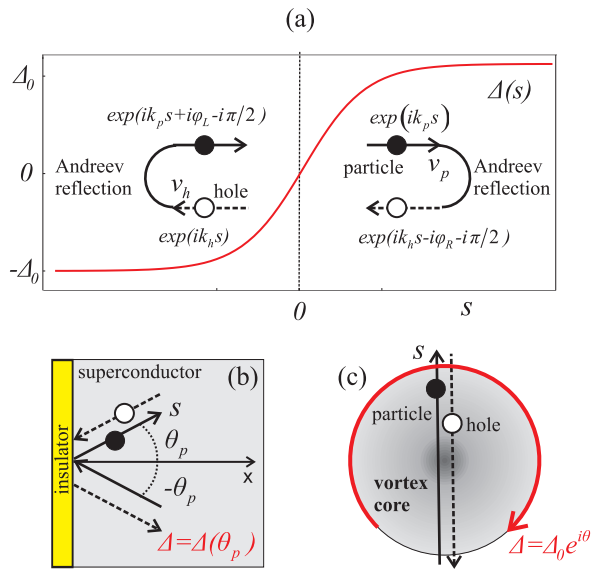}}
\caption{ \label{Fig:AndreevReflection}  Schematically shown
formation of Andreev bound states localized  (a) on domain wall,
(b) on the edge, (c) inside the vortex core. In all cases the
mechanism is the subsequent particle-hole conversions via Andreev
reflections at the opposite ends of the trajectory $s$. The
reflected particle (hole) picks up the phase of the order
parameter $\varphi_R (-\varphi_L)$ and flips the group velocity
direction ${\bm v_p} ({\bm v_h})$ as shown in the panel (a). In
general the wave vectors of particle and hole in the bulk are
slightly different $k_{p,h}=k_F\pm E/v_F$ where $k_F$ and $v_F$
are Fermi momentum and velocity, $E$ is the energy. In case if the
order parameter phase difference is $\phi_R-\phi_L=\pi$ the closed
loop can be formed even for $k_e=k_h$, that is for the zero energy
$E=0$. In cases (b,c) the phase difference appears due to the
momentum dependence of the gap function and the phase winding
around  the vortex core correspondingly. }
\end{figure}

Here we consider the gapless Majorana fermions, which appear as
Andreev bound states on the surfaces of superfluids and on
topological objects in superfluids: quantized vortices, solitons
and domain walls.  In all cases the bound states are formed
due to the subsequent Andreev reflections of particles and holes.
The key factor for the formation of ABS on the small defect with
the size of the order of coherence length is a non-trivial phase
difference of the order parameter at the opposite ends of particle
trajectory. In general it depends on the structure of the order
parameter in real and momentum space which can be rater
complicated. The possibilities for the formation of ABS are rather
diverse, several of them are shown in in
Fig.\ref{Fig:AndreevReflection}. The particularly interesting are
the case when ABS are topologically stable, which means that they
have stable zero-energy Majorana modes which cannot be eliminated
by the small perturbation of the system parameters.

General properties of fermionic spectrum in condensed matter and particle physics
are determined by topology of the ground state (vacuum). The classification schemes based on
topology \cite{Schnyder2008,Schnyder2009a,Schnyder2009b,Kitaev2009,Volovik2003,Volovik2007,Horava2005}
suggest the classes of  topological insulators, fully gapped topological superfluids/superconductors
and gapless topological media. In Refs. \cite{Volovik2003,Volovik2007,Horava2005} the classification
is based on topological properties of matrix Green's function, while the other schemes explore the properties of single
particle Hamiltonian and thus are applicable only to systems of free (non-interacting) fermions.
Among the fully gapped topological superfluids
there is the time-reversal invariant  superfluid $^3$He-B, thin films of chiral superfluid $^3$He-A
and  thin films of  the time-reversal invariant  planar phase of superfluid $^3$He.  The main signature of topologically
nontrivial vacua with the energy gap in bulk
is the existence of zero-energy edge states on the boundary, at the interface between topologically distinct
domains \cite{HasanKane2010,Xiao-LiangQi2011} and in the vortex cores \cite{SilaevVolovik2010}.
For superfluids and superconductors these are
Andreev-Majorana bound states (AMBS). These are mainly propagating fermionic quasiparticles, which have relativistic spectrum
at low energy \cite{SalomaaVolovik1988,Volovik1997,ChungZhang2009,Volovik2009a,Nagato2009,Volovik2009b}.
However, for special geometries and dimensions, the Andreev-Majorana bound state represents the isolated
non-propagating midgap state, called the Majorana zero mode (or Majorino \cite{Wilczek2014}).
It is not a fermion, since it obeys a non-Abelian exchange statistics
\cite{Ivanov2001}. This in particular occurs for the AMBS in the vortex core of chiral $p$-wave
superfluid-superconductor in 2+1 dimensions \cite{Volovik1999}.

The gapless AMBS takes place also on the surfaces, interfaces and in the vortex cores of the gapless topological media.
 Among them there are chiral superfluid $^3$He-A with Weyl points, the time-reversal invariant planar phase
with Dirac points and the time-reversal invariant polar phase with line of zeroes.
 The spectrum of Andreev-Majorana bound states there is non-relativistic and exotic: the zeroes of AMBS spectrum form Fermi arcs
\cite{Tsutsumi2011,XiangangWan2011,Burkov2011,SilaevVolovik2012} and flat bands
\cite{KopninSalomaa1991,TanakaKashiwaya1995,Ryu2002,SchnyderRyu2010,HeikkilaVolovik2011,HeikkilaKopninVolovik2011,Volovik2011,SatoTanakaYokoyama2011}.

\section{Andreev-Majorana edge states in 2+1 gapped topological superfluids}
\label{2Dgapped}

The $p$-wave superfluid $^3$He has been discovered  in 1972. However till now there is little understanding
of superfluid $^3$He films. The information on recent experiments in confined geometry can be found in  review \cite{Levitin2014}.
In thin films the competition is expected between the chiral superfluid $^3$He-A and time-reversal invariant  planar phase,
both acquiring the gap in the spectrum  in quasi-two-dimensional case due to transverse quantization.

 The fermionic spectra in both the 2D A phase and the planar phase have non-trivial
 topological properties. These topological states provide the examples of systems featuring
  generic topological phenomena. In particular the analog of integer quantum Hall effect
 exist in the 2D A phase where the internal orbital momentum of Cooper pairs plays the role
 of time reversal symmetry breaking magnetic field. In the time reversal invariant planar phase
 the quantum spin Hall effect can be realized. In close analogy with the 2d electronic systems
 the topological invariant is determined by the number of fermionic edge modes with zero energy.
  In the superfluid systems the edge zero modes are the ABS localized at the superfluid/vacuum
  boundary or at the interfaces and domain walls separating superfluid states
  with different topological properties. Below we discuss in detail the topological
   properties and ABS for the 2D A phase and the planar phase.

 \subsection{Chiral $^3$He-A film}

  The order parameter in spatially homogeneous time reversal symmetry breaking $^3$He-A phase
 is given  $\hat\Delta=\sigma_x (p_x\pm ip_y)$ where $\sigma_x$ is spin
 Pauli matrix and the $p_{x,y}$ are momentum projections to the
 anisotropy plane. Such order parameter describes the
 triplet Cooper pairs with zero spin $S_z=0$ and non-zero oribital momentum $L_z=\pm 1$ projections onto the anisotropy axis.
 The non-zero $L_z$ plays the role of internal magnetic field
 breaking the time-reversal symmetry of the systems. Confined in $zy$ plane the 2D
 state of A phase is a fully gapped system. By the analogy with
 2D electronic gase in quantized magnetic field the gapped ground states (vacua) in 2+1 or quasi 2+1 thin
films of $^3$He-A are characterized by the following  topological
invariant
\cite{So1985,IshikawaMatsuyama1986,IshikawaMatsuyama1987,Volovik1988,VolovikYakovenko1989}:
\begin{equation}
N = \frac{e_{ijk}}{24\pi^2}~
{\bf tr}\left[  \int    d^3p
~G\partial_{p_i} G^{-1}
G\partial_{p_j} G^{-1}G\partial_{p_k}  G^{-1}\right].
\label{2+1invariant}
\end{equation}
Here $G=G(p_x,p_y,\omega=ip_0)$ is the Green's function matrix,
which depends on Matsubara frequency $p_0$; the integration is
over the whole (2+1)-dimensional momentum-frequency space
$p_i=(p_x,p_y,p_0)$, or over the Brillouin zone and $p_0$ in
crystals. The expression (\ref{2+1invariant}) is the extension of 
the TKNN invariant invented by Thouless {\it et al.} to describe the topological
quantization of Hall conductance\cite{TKNN1,TKNN2}. 

\begin{figure}[hbt]
\centerline{\includegraphics[width=0.90\linewidth]{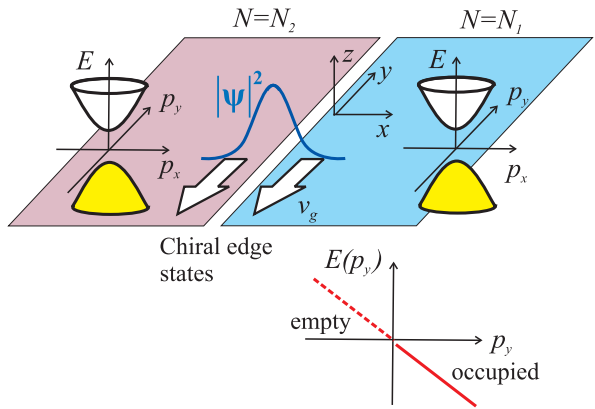}}
\caption{ \label{Fig:EdgeStates}  Schematic picture of
 the interface between two films of
chiral $p_x+ip_y$ superfluid with values $N_1$ and $N_2$ of
topological invariant (\ref{2+1invariant}). The interface contains
chiral AMBS with spectrum  $E=E(p_y)$, which move with group velocity
$v_g=dE(p_y)/dp_y$. In general the algebraic sum of branches (the
number of left moving $-$ the number of right moving fermions) is
$N_2-N_1$. On the lower panel the chiral branch of spinless AMBS
is shown given by Eq.(\ref{2DSpinPolarized}), when $N_2=1$ and
$N_1=0$. For the spinful case in Eq.(\ref{3He-A_film}) there are
two anomalous branches of spectrum of edge states $E(p_y)$ which
are degenerate over spin.  The chiral branches produce an
equilibrium mass current flowing along the interface. }
\end{figure}

The advantage of the topological approach is that one can choose for consideration  the simplest form of Green's function, which has the same topological
properties and can be obtained from the complicated one by continuous deformation. For a single layer of $^3$He-A film one can choose
 \begin{equation}
  G^{-1}=ip_0 +   \tau_3\left(\frac{p^2}{2m} -\mu\right) + c\sigma_z \left(\tau_1 p_x  + \tau_2 p_y\right).
 \label{3He-A_film}
 \end{equation}
 where $ p^2=p_x^2+p_y^2$. The Pauli matrices $\tau_{1,2,3}$ and $\sigma_{x,y,z}$ correspond to the
Bogoliubov-Nambu spin and ordinary spin of $^3$He atom
respectively;  the parameter $c$ characterizes the amplitude of
the superconducting order parameter. The  weak coupling BCS  limit
corresponds to $mc^2 \ll \mu$. In this limit one has $c=\Delta/p_F$,
where $\Delta$ is the gap in the spectrum and $p_F$ is Fermi
momentum, $p_F^2/2m=\mu$.

 It is also instructive to consider the simplified case when there is only
a single spin component, which corresponds to the
fully spin polarized $p_x+ip_y$ superfluid:
 \begin{equation}
   G^{-1}=ip_0 +   \tau_3\left(\frac{p^2}{2m} -\mu\right) + c\left(\tau_1 p_x  + \tau_2 p_y\right).
\label{2DSpinPolarized}
\end{equation}
We call this case as spinless fermions. The topological invariant (\ref{2+1invariant}) for the state in
Eq.(\ref{2DSpinPolarized}) with $\mu>0$ is $N=1$, while for the state with $\mu<0$ one has $N=0$.
 According to the bulk-surface correspondence, at the interface between these two phases there must
 be the branch of the Andreev-Majorana edge states, which crosses zero energy level
 \cite{Volovik1992,Volovik1997}, see Fig. \ref{Fig:EdgeStates}.

In the spinful case of Eq.(\ref{3He-A_film}), both spin components equally contribute to the topological invariant,
and one  has $N=2$ for $\mu>0$ and $N=0$ for $\mu<0$. Therefore  there must be two branches of the
Andreev-Majorana edge states, which cross zero energy level.
In general case the algebraic sum of anomalous branches (the number of left moving minus the
 number of right moving fermions) satisfies the index theorem, $n_L-n_R=N(x>0)-N(x<0)$.

 \subsection{Time-reversal invariant planar phase}

 Alternative to the 2D chiral A phase in thin films of
superfluid $^3$He the time-reversal invariant planar
phase~\cite{VW} can become stable. While this phase has not been
identified experimentally yet, in recent
experiments~\cite{Saunders1,Saunders2,SaundersSci13,SaundersPRL13,Levitin2014}
strong suppression of the transverse gap has been observed.

The order parameter which describes the spatially homogeneous time
reversal invariant planar phase has the form $\hat\Delta=p_y + i
\sigma_z p_x$. In this phase, the order parameters is anisotropic
and vanishes for the ${\bf p} \parallel {\bf z}$ direction,
transverse to the film. Nevertheless, confined in 2D when $p_z=0$
this system is gapful.

\begin{figure}[hbt]
\centerline{\includegraphics[width=1.0\linewidth]{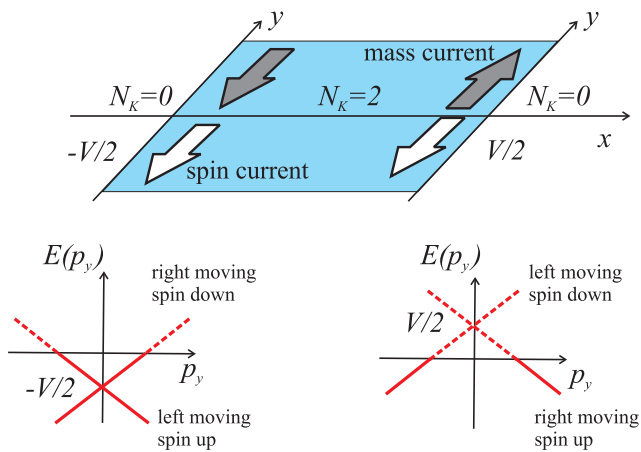}}
\caption{ \label{Fig:SpinChargeHall}   An illustration of the
intrinsic spin-current quantum Hall effect  due to
Andreev-Majorana edge states in the stripe of the planar phase
film with topological invariant $N_K=2$ in Eq. (\ref{N2+1prime}).
As distinct from $^3$He-A in Fig. \ref{Fig:EdgeStates}, the
anomalous branches with different spin projections have opposite
slopes. This gives rise to the quantized spin Hall effect without
magnetic field, instead of the quantized Hall effect in $^3$He-A
film \cite{VolovikYakovenko1989,Volovik1992b}. }
\end{figure}

Being time-reversal invariant the planar phase has zero
topological invariant of the type given by
Eq.(\ref{2+1invariant}). However it has an extra discrete
symmetry, namely a combination of a $\pi$ spin rotation around
 $z$-axis followed by a $\pi/2$ phase rotation. This modifies the
topological classification, adding extra $\mathbb{Z}$ topological
invariant obtained by Volovik and Yakovenko in
Ref.~\onlinecite{VolovikYakovenko1989}. This invariant gives rise
to the intrinsic spin-Hall effect illustrated in Fig. \ref{Fig:SpinChargeHall}.

An extra motivation to study this particular case of the planar
phase is that it can be considered as a corner stone for a
dimensional reduction scheme which can be applied to general
class-DIII topological superconductors. In the next section we
discuss that the topological properties of a 3D system and an
embedded (2+1)D system, which exist in any time-reversal invariant
cross section of the momentum space, are connected. As an
application of such a reduction we derive a generalized index
theorem for 3D topological superconductors, which provides an
example of the bulk-boundary correspondence in odd spatial
dimensions.

 In the single layer case, the simplest expression for the  planar phase Green's function  $G(p_0,p_x,p_y)$
  is determined by
 \begin{equation}
  G^{-1}= i p_0 +  \tau_3\left(\frac{p^2}{2m} -\mu\right) +c \tau_1 (\sigma_x p_x  + \sigma_yp_y)\,.
 \label{planar_state}
 \end{equation}

This phase is symmetric under time reversal. The two spin components have opposite chiralities, as
can be seen from identity
 \begin{equation}
  \sigma_x p_x  + \sigma_yp_y=\frac{1}{2}(\sigma_x + i\sigma_y)(p_x-ip_y) +
\frac{1}{2}(\sigma_x - i\sigma_y)(p_x+ip_y) \,.
 \label{planar_components}
 \end{equation}
 That is why the contributions of the two spin components to the topological invariant (\ref{2+1invariant}) cancel each other, $N=0$. But the planar phase
is still topologically non-trivial because of the discrete $Z_2$-symmetry between the two spin components
in Eq.(\ref{planar_components}).  Due to this symmetry the matrix  $K=\tau_3\sigma_z$ commutes with the Green's function, which allows us to introduce the symmetry protected topological invariant\cite{VolovikYakovenko1989,Volovik1992b}:
\begin{equation}
 N_K= \frac{e_{ijk}}{{24\pi^2}} ~
{\bf tr}\left[ K \int    d^3p
~G\partial_{p_i} G^{-1}
G\partial_{p_j} G^{-1}G\partial_{p_k}  G^{-1}\right].
\label{N2+1prime}
\end{equation}
This invariant is robust to deformations, if the deformations are $K$-symmetric. For the state
(\ref{planar_state}) with $\mu>0$ one has $N_K=2$. For the general case of the quasi 2D  film with multiple layers of the planar phase, the invariant $N_K$ belongs to the group $Z$.  The magnetic solid state analog of the planar phase is the 2D time reversal invariant topological insulator, which experiences the quantum spin Hall effect without external magnetic field \cite{HasanKane2010}.

Fig. \ref{Fig:SpinChargeHall} demonstrates    Andreev-Majorana edge states on two boundaries of  the stripe of the single layer of planar phase film. As distinct from $^3$He-A in Fig. \ref{Fig:EdgeStates}, the anomalous branches with different spin projections are not degenerate: they have opposite slopes, which corresponds to the zero value of the invariant $N=0$ in Eq.(\ref{2+1invariant}).  In case of superconductor with planar phase symmetry, the invariant $N_K$ determines quantization of spin Hall effect. In applied voltage $V$ the spectra on two boundaries  shift in opposite directions, changing the population of branches. This produces the imbalance in the spin currents carried by edge states on two boundaries, giving rise to the non-zero total spin-current $J_x^z$ (current of $z$-projection of spin along $x$-axis). This is in the origin of quantized spin Hall effect in the absence of magnetic field
\cite{VolovikYakovenko1989,Volovik1989,Volovik1992b}:
\begin{equation}
J_x^z= \sigma_{xy}^{\rm spin}E_y~~,~~ \sigma_{xy}^{\rm spin}=\frac{N_K}{4\pi}\,.
\label{SCHE}
\end{equation}
 In this time reversal invariant system the electric current QHE is absent. The topological charge $N$ in Eq.(\ref{2+1invariant}), which determines quantization of the Hall conductance in the absence of magnetic field \cite{Volovik1988}, is $N=0$, and the currents of different spin populations cancel each other.

The mass and spin currents carried by AM edge state in $p$-wave superfluids have been considered in Refs. \cite{Sauls2011,WuSauls2013}.

\section{AMBS on surface of 3+1 gapped topological superfluid}
\label{AMBSgapped}

 Fully gapped $3+1$ fermionic systems - topological
insulators and topological superconductors - are now under
extensive investigation. The interest to such systems is revived
after identification of topological insulators in several
compounds \cite{HasanKane2010}.

These systems are characterized by the gapless fermionic states on
the boundary of the bulk insulator or at the interface between
different states of the insulator. Historically, the topological
insulators with fermionic zero modes at the interface have been
introduced in works \cite{Volkov}.  An example of the fully gapped
topological superfluids is the B phase of superfluid $^3$He.
Much attention has been devoted to the investigation of
bound fermion states on surface of $^3$He-B. The presence of Andreev-Majorana
surface states in $^3$He-B
can be probed through anomalous transverse sound attenuation
\cite{NagaiImpedance,ImpedanceExp1,ImpedanceExp2,ImpedanceExp3},
and surface specific heat measurements \cite{SpecificHeat,Bunkov2014}.
These AM bound states are supported by the non-zero value of the
topological invariant in $^3$He-B \cite{Volovik2009b} and have two -
dimensional relativistic massless Dirac
spectrum
\cite{Nagato2009,SalomaaVolovik1988,Volovik2009a,ChungZhang2009,Tsutsumi2011}.

 \subsection{$^3$He-B edge states from bulk topology}

 \begin{figure}[top]
\centerline{\includegraphics[width=0.9\linewidth]{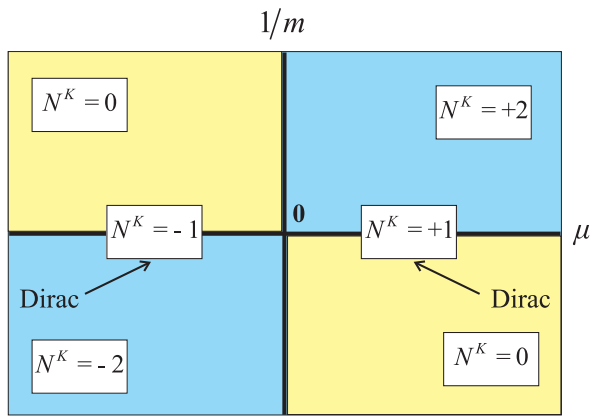}}
  \caption{\label{QPT}  Phase diagram of topological states of triplet superfluid of $^3$He-B type in
  equation \eqref{eq:HamiltonianHeB} in the plane $(\mu,1/m)$. States on the line
  $1/m=0$ correspond to the  Dirac vacua, which Hamiltonian is non-compact. Topological charge of the Dirac fermions
  is intermediate between charges of compact $^3$He-B states.
   The line $\mu=0$ marks topological quantum phase transition, which occurs between the weak coupling $^3$He-B
   (with $\mu>0$, $m>0$ and topological charge $N_K=2$) and the strong coupling $^3$He-B (with $\mu<0$, $m>0$ and $N_K=0$).
   This transition is topologically equivalent to quantum phase transition between Dirac vacua with opposite mass parameter
 $M=\pm |\mu|$.  The gap in the spectrum becomes zero at this transition.
The line $1/m=0$ separates the states with different asymptotic
behavior of the Hamiltonian at infinity: $H({\bf p}) \rightarrow
\pm \tau_3 p^2/2m$. The transition across this line  occurs
without closing the gap.
 }
\end{figure}

The  topological
superfluid/superconductor of  the $^3$He-B type are described by topological invariant $N_K$, which is protected by symmetry:
\begin{equation}
N_K = {e_{ijk}\over{24\pi^2}} ~
{\bf tr}\left[K \int   d^3p
~H^{-1}\partial_{p_i}H
H^{-1}\partial_{p_j}H
H^{-1}\partial_{p_k}H
\right]\,.
\label{3DTopInvariant_tau}
\end{equation}
Here $H({\bf p})$ is the Hamiltonian, or in case of interacting system the inverse Green's function at zero frequency $H({\bf p})=G^{-1}(\omega=0,{\bf p})$, and  $K$ is matrix which  commutes or anti-commutes with $H({\bf p})$.

The proper model Hamiltonian which has the same topological properties as superfluids/superconductors of  the $^3$He-B class is the following :
\begin{equation}
H=\left(\frac{p^2}{2m}-\mu\right) \tau_3-
   c \tau_1 {\mbox{\boldmath$\sigma$}} \cdot{\bf p} \,,
\label{eq:HamiltonianHeB}
\end{equation}
where $\tau_i$ and $\sigma_i$ are again the Pauli matrices of Bogolyubov-Nambu spin  and nuclear spin correspondingly.
The symmetry $K$, which enters the topological invariant $N_K$ in \eqref{3DTopInvariant_tau},
  is represented by the $\tau_2$ matrix, which anti-commutes with the Hamiltonian: it is the combination of time reversal
   and particle-hole symmetries of $^3$He-B.
   In the limit $1/m=0$, Eq.(\ref{eq:HamiltonianHeB}) transforms to the Dirac Hamiltonian, where
the parameter $c$ serves as the speed of light, while $^3$He-B lives in the opposite limit  $mc^2\ll \mu$.
The topological phase diagram in the plane of parameters $\mu, 1/m$  is in Fig. \ref{QPT}.

   The mechanism of Andreev-Majorana  bound states formation at the edge of
   $^3$He-B  is clear from the Hamiltonian (\ref{eq:HamiltonianHeB}).
   Let us consider the boundary plane at $x=0$ as shown
   schematically in the Fig.(\ref{Fig:ReductionBtoPlanar}). Then
   upon the normal reflection of particles and holes from the
   boundary some components of the gap function in (\ref{eq:HamiltonianHeB})
   change sign. Therefore one obtains a non-zero phase of the gap along
   the effective trajectory as shown in
   Fig.(\ref{Fig:AndreevReflection}). In particular for the
   trajectories normal to the boundary $p_{x,y}=0$ the overall gap
   function changes the sign leading to the formation of the
   zero-energy state localized at the boundary.

   However this is not the whole story. Indeed if one formally
   assumes that the Hamiltonian may have either negative effective mass, $m<0$,
   or negative chemical potential, $\mu<0$, the exact solution of the spectral
   problem yields no zero-energy states as will be discussed
   below. The hint to the topological origin of the AMBS in $^3$He-B can
   be obtained from the topological phase diagram in Fig. \ref{QPT}, which
   demonstrates that the system undergoes a
   topological quantum phase transitions as one changes the sign of
   chemical potential $\mu$ or effective mass $m$.

The domain wall, which separates the states with different values of $N_K$, should
contain the zero energy states -- the Andreev-Majorana zero modes.

 \subsection{$^3$He-B edge states from topology of planar phase}

 \begin{figure}[hbt!]
 \centerline{\includegraphics[width=1.0\linewidth]{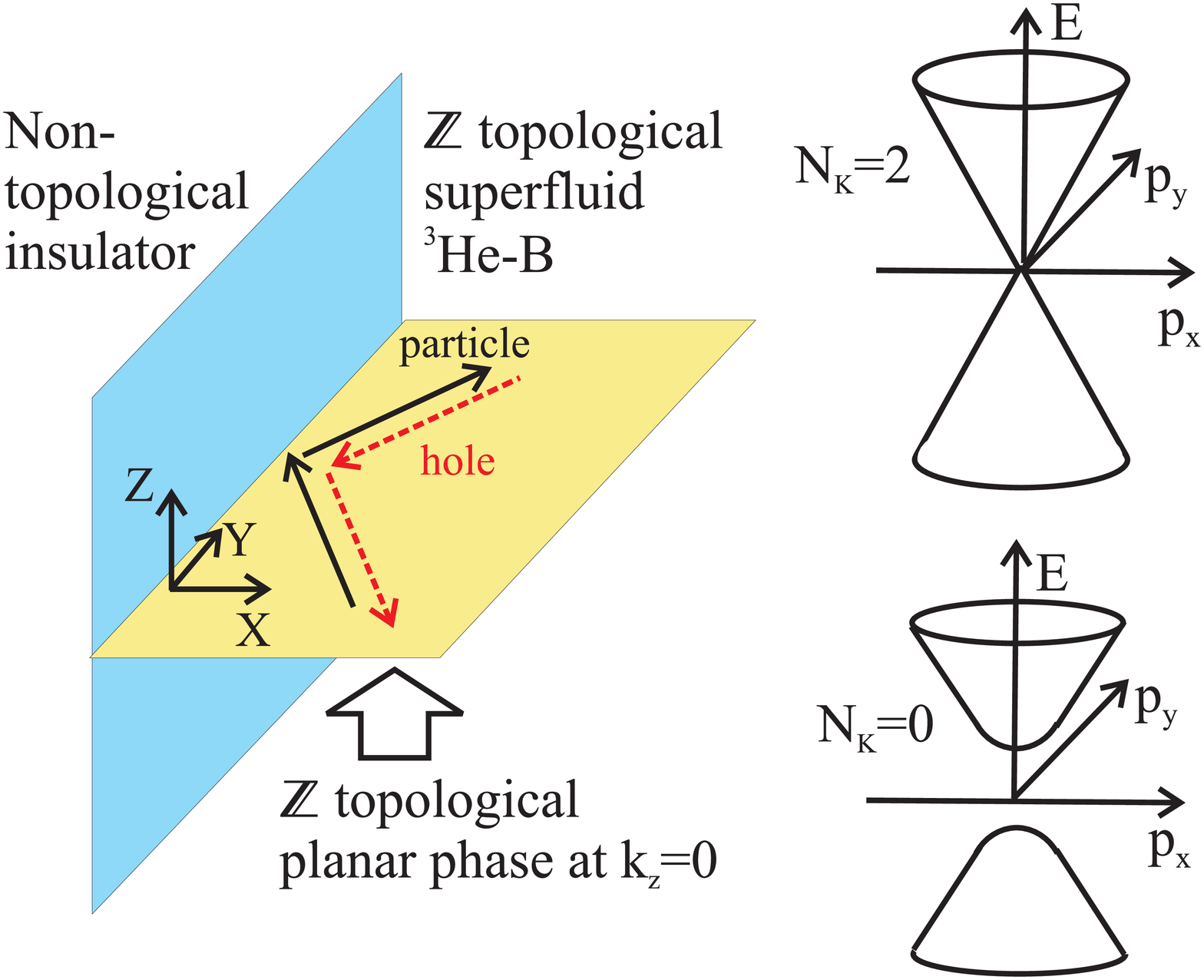}}
   \caption{\label{Fig:ReductionBtoPlanar} (Color online)
  Dimensional reduction of the surface-states spectral problem in
  3D to that in the time-reversal invariant cross section of momentum space $k_z=0$.
   Reduction from the
 $\Z$ topological superfluid $^3$He-B results in the $\Z$ topological planar phase at $k_z=0$.
   }
 \end{figure}

To prove the existence of the Andreev-Majorana bound states on the
surface of $^3$He-B or at the interface one can use a dimensional
reduction. Let us assume that the boundary plane is at $x=0$, so that
the conserved longitudinal momentum projections are $k_{z,y}$. To
find the complete spectrum of bound states $E_b=E_b(k_y,k_z)$ it
is enough to consider a set of 2D spectral problems for the cross
sections of momentum space
 \begin{equation}\label{Eq:Line}
  k_y \cos\theta + k_z \sin\theta=0 \,,
 \end{equation}
where $2\pi>\theta\geq 0$.

An example of such a dimensional reduction to the plane $k_z=0$ is
shown in Fig.\ref{Fig:ReductionBtoPlanar}.  The $2+1$ Hamiltonian
in this cross section reduced from the $3+1$ phase exactly
coincides with the Hamiltonian of the planar phase. Therefore it
is classified by the integer-valued topological invariant $N_K$ in
Eq.(\ref{N2+1prime}), which can be shown to coincide with
 the topological invariant $N_K$ of the parent 3D $^3$He-B phase in Eq.(\ref{3DTopInvariant_tau}).
The topologically protected Andreev-Majorana states in $^3$He-B are thus related to the topologically
protected edge states in the $2+1$ planar phase, see details in Ref. \cite{Makhlin2014}.

 \subsection{Evolution of edge state at non-topological QPT}

Let us consider the spectrum of Andreev-Majorana fermions using the simplest model of
the interface between the superfluid and  the vacuum, in which the
Hamiltonian (\ref{eq:HamiltonianHeB}) changes abruptly at the boundary, with the boundary condition
 $\psi (z=0)=0$.

\begin{figure}[hbt]
\centerline{\includegraphics[width=0.90\linewidth]{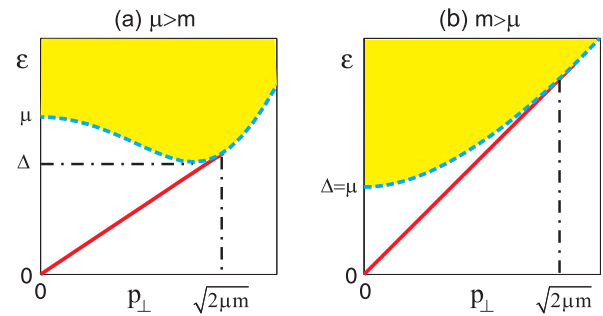}}
\caption{\label{Fig:Spectrum}   Spectrum of Andreev-Majorana
fermions, localized states on the surface of topological
superfluid/superconductor of the $^3$He-B class (red solid lines)
for (a) $\mu>m>0$ and (b) $m>\mu$. The spectrum of bound states
terminates when it merges with continuous spectrum in bulk (yellow
color), whose border is shown by blue dashed line. The AM bound
states exist for $p_\perp^2<2m\mu$. }
\end{figure}

At low energies $|E|\ll \Delta$  their spectrum is helical
spectrum, being described by the Hamiltonian  $H_{\rm
AM}=c(\sigma_y p_x - \sigma_x p_y)$ \cite{ChungZhang2009}.
Interestingly an exact solution of the spectral problem
  demonstrates \cite{SilaevVolovikJETPL2012} that the linear spectrum of AMBS exist up to
  the merging point with the continuous spectrum of delocalized states.

For $m>0$ the exact spectrum of Andreev-Majorana fermions $E=\pm
p_\perp$ is shown by the red solid line in Fig. \ref{Fig:Spectrum}
for $E>0$.
The bound states are confined to the
region $|p_\perp|<\sqrt{2m\mu}$. They disappear when their
spectrum merges with the continuous spectrum in bulk. The edge of
continuous spectrum is shown by blue dashed line in
Fig. \ref{Fig:Spectrum}. If $mc^2>\mu$  the minimum of the bulk energy spectrum
 increases
monotonically with momentum $p_\perp$, therefore the bulk gap is
\begin{equation}
\Delta=\mu~~,~~mc^2>\mu\,.
\label{gap1}
\end{equation}
If $\mu>mc^2$ the minimum of the bulk energy is non-monotonic function of
$p_\perp$ having the minimum at
$p^{min}_\perp=\sqrt{2m(\mu-mc^2)}$ where the bulk gap is
\begin{equation}
\Delta=\sqrt{mc^2(2\mu -mc^2)}~~,~~0<mc^2<\mu\,.
\label{gap2}
\end{equation}
The line $mc^2=\mu$ marks the non-topological quantum phase transition -- the momentum space analog of
the Higgs transition \cite{Volovik2007}, when the Mexican hat  potential as function of $p_\perp$ emerges for $\mu > mc^2$ .

 \subsection{Evolution of edge state at topological QPT}

Let us first consider the behavior of the spectrum of Majorana
fermions at the topological transition at which $m$ crosses zero.
When $m$ approaches zero, $m\rightarrow 0$, the region of momenta
where bound states exist shrinks  and finally for $m<0$, i.e. in
the topologically trivial superfluid, no bound states exist any
more. Simultaneously the gap in bulk, which at small $m$ is
$\Delta\approx \sqrt{2mc^2\mu}$ according to Eq. \eqref{gap2},
 decreases with decreasing $m$
and nullifies at $m=0$. This corresponds to the conventional scenario of the topological quantum phase transition,
 when at the phase boundary between the two gapped states with different topological
numbers the gap is closed. The same happens at the TQPT occurring when $\mu$ crosses zero
(see phase diagram in Fig. \ref{QPT}).

Now let us consider what happens with bound states in the case if
the TQPT occurs in the opposite limit, when $m$ changes sign via
infinity, i.e. when $1/m$ crosses zero. This topological
transition occurs without closing of the gap. In this case the
bound states formally exist for all $p_x$ even in the limit $1/m
\rightarrow 0$. However, in this limit the ultraviolet divergence
takes place: the characteristic length scale of the wave function
of the bound state  $L \propto \hbar/mc \rightarrow 0$. So, if the
TQPT from topologically non-trivial to the trivial  insulator (or
superconductor)
 occurs without closing the gap, the gapless spectrum of surface states disappears by escaping via ultraviolet. This limit corresponds to
  formation of zero of the Green's function, $G=1/(i\omega -H)\rightarrow 0$.
 Such scenario is impossible
 in the models with the bounded Hamiltonian \cite{Gurarie2011,Essin2011}, which takes place  in approximation
  of finite number of crystal bands.

On the other hand Green's function zeroes can occur due to the
particle interactions.   As was found in Ref.
\cite{FidkowskiKitaev2010}, classifications of interacting and
non-interacting fermionic systems do not necessarily coincide.
This is related to zeroes of the Green's function, which according
to Ref.  \cite{Volovik2007} contribute to topology alongside with
the poles. Due to zeroes the integer topological charge of the
interacting system can be changed without closing the energy gap,
and it is suggested that this may lead to the occurrence of
topological insulators with no fermion zero modes on the interface
\cite{Gurarie2011,Essin2011}.

  That is why we expect that the same scenario with escape
to the ultraviolet takes place for the interacting systems:  if
due to zeroes in the Green's function the TQPT in bulk occurs
without closing the gap, the spectrum of edge states will
nevertheless change at the TQPT, and  this change occurs via the
ultraviolet.

Finally let us mention, that the magnetic field violates time reversal symmetry, which generically 
leads to the finite gap (mass) in the spectrum of AM fermions on the surface. At particular orientation
of the magnetic field there is still the $Z_2$ dsicrete symmetry, which supports gapless AMBS
\cite{MizushimaSatoMachida2012,Mizushima2012}. This symmetry is spontaneously  broken at some critical 
value of magnetic feild, above which the AM fermions become massive. The surface of $^3$He-B with massive AM bound states represents the $2+1$ topological "insulator": it is described by the  topological invariant in Eq.(\ref{2+1invariant}). 
The line on the surface, which separates the surface domains with different values of this topological invariant, contains $1 + 1$ gapless AM fermions  \cite{Volovik2010b}.

\section{AMBS on surface of 3+1 Weyl superfluid. Fermi arc}
\label{FermiArc}

Now we move to the AM bound states which appear as edge and vortex states in the
gapless topological systems. Here the zeroes in bulk lead to the extended zeroes on the surfaces,
interfaces and vortex cores. We start with point zeroes --  Weyl points -- in chiral superfluids, which
produce the lines of zeroes (Fermi arc) on the surface, and the flat band in the vortex core.

\subsection{Andreev-Majorana Fermi arc on the boundary of Weyl superfluid}
\label{FermiArcBoundary}

The topological origin of AM bound states in $3+1$ chiral superfluids can be viewed by extension of the
topology of the $2+1$ chiral system in Sec. \ref{2Dgapped} to the $3+1$ case.
Let us consider for simplicity the spinless fermions, or which is the same, the fermions with a given spin polarization.
Then the Green's function in Eq.(\ref{3He-A_film}) extended to $3+1$ case is:
 \begin{equation}
  G^{-1}({\bf p}, p_0)=ip_0 +   \tau_3\left(\frac{p^2}{2m} -\mu\right) +  c\left(\tau_1 p_x  + \tau_2 p_y\right).
\label{3He-A_bulk}
\end{equation}
where ${\bf p} = (p_x,p_y,p_z)$. Let us consider $p_z$ as parameter of the $2+1$ system. Then for each $p_z$, except for
$p_z=\pm p_F$, this Green's function describes the fully gapped
$2+1$ system -- the "insulator", which is characterized by
topological invariant in Eq.(\ref{2+1invariant}):
 \begin{equation}
 \begin{split}
 &  N(p_z)
 \\
 & =\frac{1}{4\pi^2}~ {\bf tr}\left[  \int    dp_xdp_ydp_0
 ~G\partial_{p_x} G^{-1} G\partial_{p_y} G^{-1}G\partial_{p_0}
 G^{-1}\right]\,.
 \end{split}
 \label{N2+1}
 \end{equation}

This insulator is topological for $|p_z| < p_F$, where $N(|p_z|<p_F)=1$, 
and  is topologically trivial  for $|p_z| > p_F$, where $N(|p_z|>p_F)=0$.

\begin{figure}[hbt]
\centerline{\includegraphics[width=0.90\linewidth]{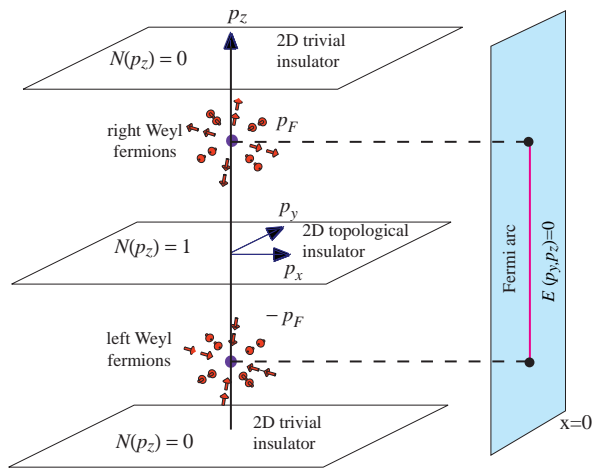}}
\caption{Line of Andreev-Majorana bound states on surface of chiral superfluid with Weyl points.
This line represents the 1D Fermi surface, which separates the edge states with positive and negative energies (see also Fig. \ref{Arc}). However, as distinct from conventional Fermi surfaces, this Fermi surface has end points, and thus is called the 
The end points of the Fermi arc are determined by projections of the bulk Weyl points to the surface.}
\label{ArcBoundary}
\end{figure}

At $p_z=\pm p_F$, the invariant (\ref{N2+1}) is not determined,
since the corresponding $2+1$ system is gapless. The bulk $3+1$
superfluid $^3$He-A has two points in the spectrum ${\bf
p}_{\pm}=(0,0, \pm p_F)$, where energy is zero, see Fig.
(\ref{ArcBoundary}). These nodes in the spectrum are topologically
protected, since they represent the monopoles in the Berry phase
in momentum space and are characterized by the topological
invariant in Eq.(\ref{2+1invariant}), where the integration now is
over the 3D sphere around the Weyl point in the $3+1$ space
$(p_0,p_x,p_y,p_z)$ \cite{Volovik2003}. In the vicinity of these
points the fermionic quasiparticles  behave as chiral (left-handed
and right-handed) Weyl fermions in particle physics. That is why
such nodes are called  the Weyl points. Arrows in Fig.
(\ref{ArcBoundary}) show the direction of the effective spin of
the Weyl fermion. This spin is parallel to
 ${\bf p}-{\bf p}_+$ in the vicinity of  ${\bf p}_+$, which means that the fermions living there are right-handed.
 For the left-handed fermions near  ${\bf p}_-$, their effective spin  is anti-parallel to
 ${\bf p}-{\bf p}_-$.

\begin{figure*}
\includegraphics[width=0.8\textwidth]{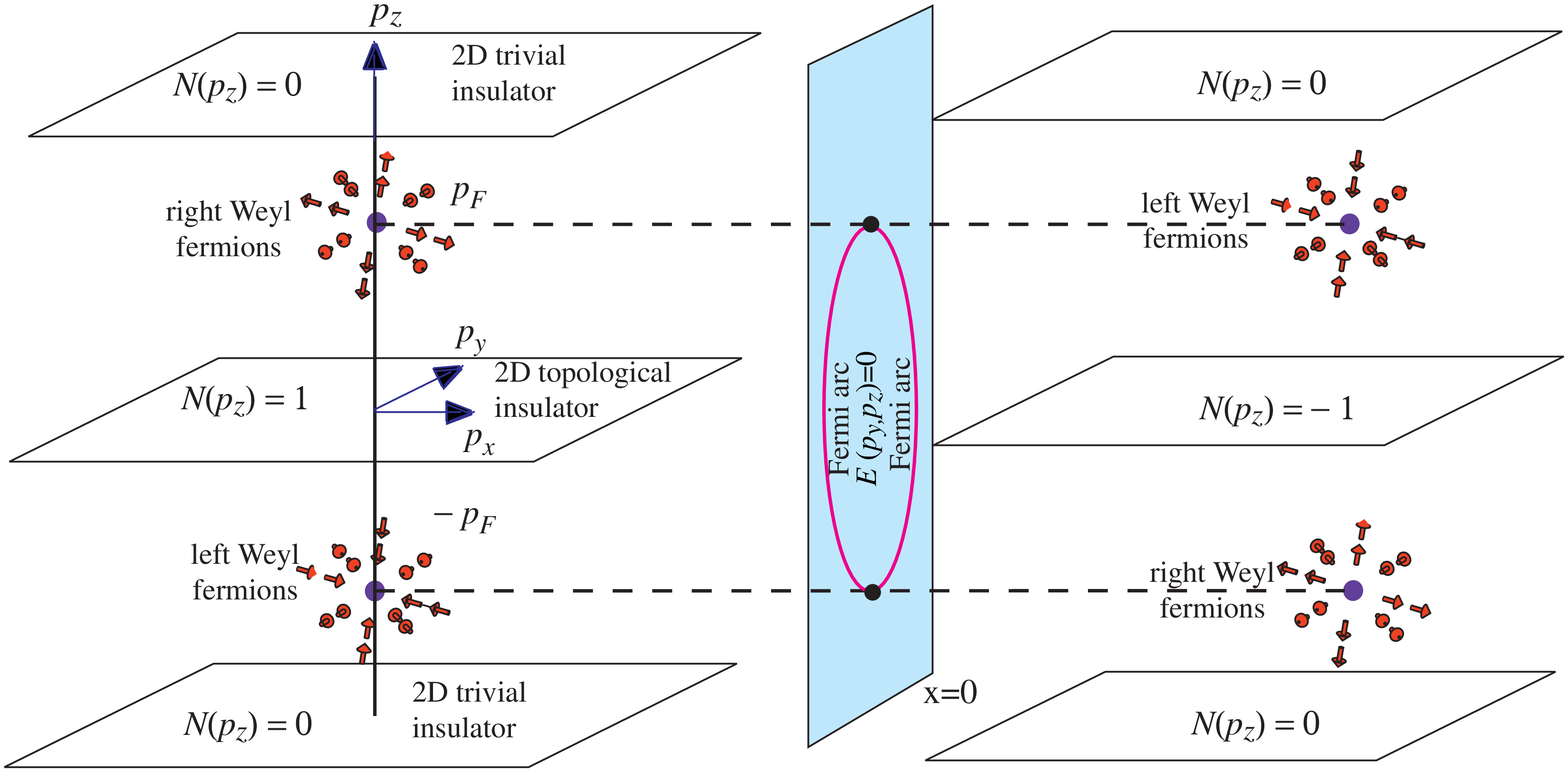}
\caption{Topology of Andreev bound states on $\hat{\bf l}$ soliton \cite{Ho1984}.
Momentum space topology of Weyl points in bulk $^3$He-A on two sides of the soliton
prescribes existence of Fermi arcs in the spectrum
of the Andreev bound states in the soliton or at the interface between the bulk states with different
positions of Weyl points. In the considered case the Weyl points on two sides of the interface have the same
positions in momentum space, but the opposite chiralities. As a result, the $2+1$ topological
insulators have opposite
topological invariants. $N(p_z=0)=\pm 1$. This leads
to two Fermi arcs terminating on the projections
of the Weyl points on the soliton/interface plane according to
the index theorem $n({\rm right}) - n({\rm left})=2$.}
\label{FermiArcSolitonFig}.
\end{figure*}

  According to the bulk-surface correspondence, at each $p_z$, for which $N(p_z)=1$, there should be  one branch of Andreev-Majorana edge states
  which cross zero energy level, see  Fig. \ref{Fig:EdgeStates}.
 As a result one has the line of zero energy states in the range $-p_F<p_z <p_F$.
 This line represents the Fermi surface (Fermi line) in the two-dimensional momentum space of bound states. As the conventional
  Fermi surface, it  separates the positive and negative energy levels But as distinct from the conventional  Fermi surface, this Fermi surface is not closed. It has two end points, and this is the reason why this line is called the Fermi arc.

 The end points of the Fermi arc coincide with the projection of the Weyl points to the surface.
 This is the consequence of bulk surface correspondence in the Weyl systems \cite{XiangangWan2011}. For the arbitrary direction of the surface with the angle $\lambda$ between the normal to the surface and the axis $z$, the  Fermi arc is concentrated
in the range of momenta  $-p_F \sin\lambda<p_z <p_F \sin\lambda$.
Note that in $^3$He-A the boundary conditions require $\lambda=0$.

In crystals, the Weyl points can be moved to the boundaries of the Brillouin zone, where they annihilate each other. As a result one obtains the chiral $3+1$ topological insulator or the fully gapped chiral topological superconductor. Since $N(p_z)=1$ for all $p_z$ the topological Fermi arc on the boundaries, transforms to the closed topological Fermi surface.

\begin{figure}[hbt]
\centerline{\includegraphics[width=0.90\linewidth]{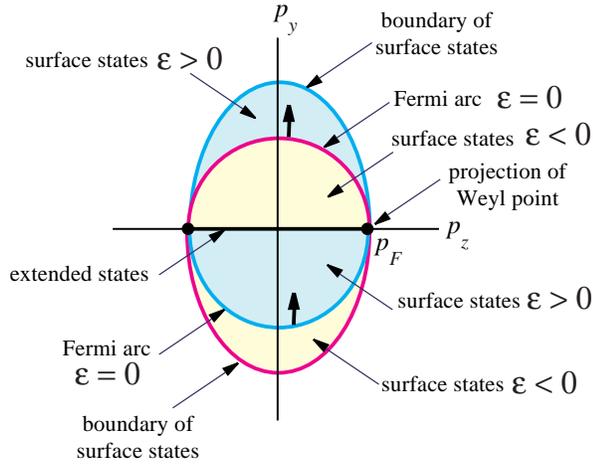}}
\caption{
The spectrum of bound states with  two Fermi arcs $\epsilon(p_y,p_z)=0$.
Thick arrows show directions of Fermi velocity at these Fermi arcs. At $p_z=0$ the velocity is in the same direction,
$v_y>0$, which demonstrates that both Fermi arcs
have the same topological charge $N=+1$, which together  satisfy the index theorem  
$n({\rm right}) - n({\rm left})=2$,
in agreement with momentum space topology
of Weyl points in bulk $^3$He-A on two sides of the soliton in Fig.~\ref{FermiArcSolitonFig}.
This leads to discontinuity in the spectrum of bound states  at $p_y=0$, where the spectrum merges
with the
bulk spectrum.  }
\label{Arc}
\end{figure}

\subsection{Andreev-Majorana Fermi arcs on soliton and domain wall}
\label{FermiArcSoliton}

The similar Fermi arcs appear on the  domain walls or solitons separating the chiral phases with opposite
chiralities. One has $N(|p_z| < p_F) =+1$ on one side of the soliton/wall and
$N(|p_z| < p_F) =-1$ on the other side. According to the
index theorem \cite{Volovik1992,Volovik2003}, the difference
between these two values determines the number of the zero modes
at the interface between the 2+1 topological insulators for each
$|p_z| < p_F$. As a result the domain wall and soliton contain two Fermi arcs instead of single Fermi arc on boundary,
see Fig. \ref{FermiArcSolitonFig}.

Fermi arc on the domain walls in $^3$He-A \cite{SalomaaVolovik1989} has been considered
in Refs.   \onlinecite{Nakahara,SilaevVolovik2012}.

Fig. \ref{Arc} includes also the bound states  with non-zero energy and demonstrates that the Fermi arc
does represent the piece of the Fermi surface, which separates the positive and negative energy levels.

\section{Topological superfluids with lines of zeroes. AM surface flat band.}
\label{PolarFlatBand}

The zero-dimensional point nodes in bulk (the Weyl points)  give rise to the one-dimensional nodes (lines)
 in the spectrum of AMBS. In the same manner,
the 1D nodal lines in bulk give rise to the 2D manifolds of AM
bound states with zero energy, see Fig. \ref{PolarPhase.eps}. Let us consider the topological
origin of such dispersionless spectrum --  the flat band -- on
example of the polar phase of triplet superfluid/superconductor
\cite{HeikkilaVolovik2011}.

\subsection{Flat band of AM modes on surface of polar phase}

The Hamiltonian for the polar phase is
\begin{equation}
H=\left(\frac{p^2}{2m}-\mu\right) \tau_3-
   c \tau_1  \sigma_z p_z \,.
\label{eq:HamiltonianPolar}
\end{equation}
This superconductor obeys the time reversal and space inversion symmetry, and it has a line of zeroes in the form of a ring.

\begin{figure}[h]
\centering
\includegraphics[width=0.8\linewidth]{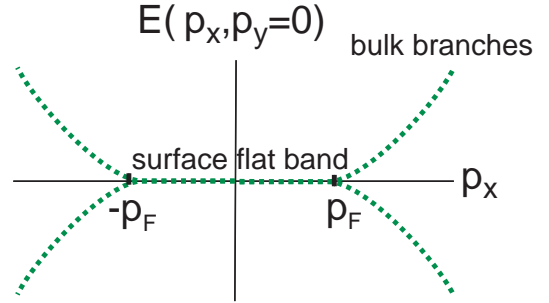}
\caption{
Spectrum of AM modes on the surface of polar phase. These modes form the 2D flat band:
all the states with $p_x^2+p_y^2< p_F^2$ have zero energy. The spectrum is shown for $p_y=0$.
 }
\label{fig:RedShift}
\end{figure}

\begin{figure}[h]
\centering
\includegraphics[width=1.0\linewidth]{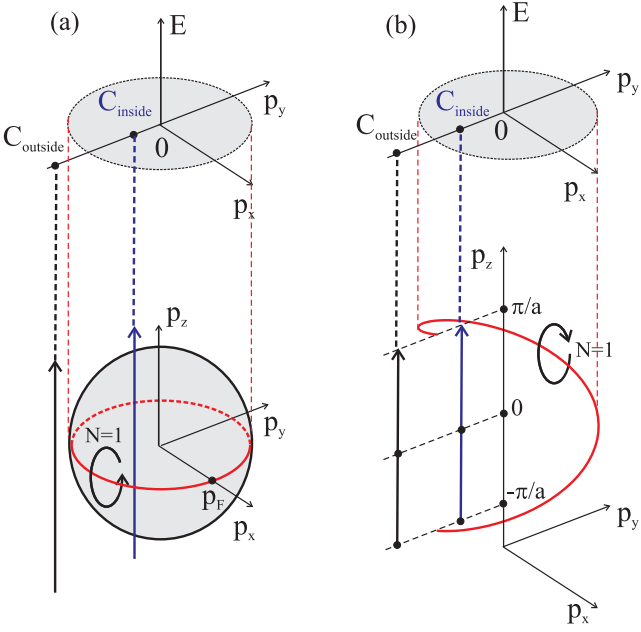}
\caption{ Topologically nontrivial nodal lines generate
topologically protected flat bands on the surface: (a) closed
equatorial line of zeroes in the polar phase; (b) spiral of zeroes
in the multilayered graphene is also the closed line. Projection
of the line on the surface determines boundary of flat band. Let
us fix $(p_x,p_y)$. If for a given $(p_x,p_y)$ the energy
$E(p_x,p_y,p_z)$ is nonzero for any $p_z$, then the Green's
function $G(\omega,p_z)_{p_x,p_y}$ describes the 1D fully gapped
system -- "insulator". At each $(p_x,p_y)$ inside the projection
of the line to the surface, this insulator is topological, since
it is described by non-zero topological invariant
(\ref{InvariantForLine2}). Thus for such $(p_x,p_y)$ there is the
gapless edge state on the surface. The manifold of these
zero-energy edge state inside the projection forms the flat band.
} \label{fig:spiral}
\end{figure}

For simplicity we consider the spinless fermions, or which is the same, the fully spin polarized fermions,
whose Hamiltonain is
\begin{equation}
H=\left(\frac{p^2}{2m}-\mu\right) \tau_3- c \tau_1  p_z \,.
\label{eq:HamiltonianPolarPolarized}
\end{equation}
The spectrum of such fermions has the nodal line -- the ring $p_x^2 + p_y^2=p_F^2$, $p_z=0$.
The stability of this nodal line is determined by the   topological invariant protected by symmetry
  \begin{equation}
N_K= \frac{1}{4\pi i} ~{\rm tr} ~K\oint_C dl ~   H^{-1}\nabla_l H \,.
\label{InvariantForLine}
\end{equation}
Here the integral is along the loop $C$ around the nodal line
in momentum space, see Fig. \ref{fig:spiral}; and the matrix $K=\tau_2$ anticommutes with the Hamiltonian.
The  winding number around the element of the nodal line is $N_K=1$.

Now let us consider the momentum ${\bf p}_\perp$ as a parameter of the $1+1$ system, then for
$|{\bf p}_\perp|\neq p_F$ the system represents the fully gapped  state -- the $1+1$  insulators.
This insulator can be described by the same invariant as in Eq.(\ref{InvariantForLine}) with
 the contour of integration chosen parallel to $p_z$. Since at $p_z\rightarrow  \pm\infty$  the Hamiltonian tends to the
 same limit, points $p_z=\pm \infty$ are equivalent, and the line $-\infty < p_z<\infty$ forms the closed loop. That is why the integral
 \begin{equation}
N_K({\bf p}_\perp)=\frac{1}{4\pi i} ~{\rm tr} ~K\int_{-\infty}^{+\infty} dp_z ~  H^{-1}\nabla_{p_z} H \,,
\label{InvariantForLine2}
\end{equation}
is integer-valued.

Topological invariant $N({\bf p}_\perp)$ in (\ref{InvariantForLine2}) determines the property of the surface bound states of
the $1+1$ system at each ${\bf p}_\perp$. Due to the bulk-edge correspondence, the topological 1D insulator must have the surface
state with exactly zero energy. Since such states exist for any parameter within the circle  $|{\bf p}_\perp|=p_F$, one obtains
the flat band of AM modes in Fig. \ref{fig:spiral} {\it left} -- the continuum of the self conjugate bound states with exactly
zero energy, $E(|{\bf p}_\perp|<p_F)=0$, which are protected by topology. Such modes do not exist for parameters
$|{\bf p}_\perp|>p_F$, for which the $1+1$ superfluid is non-topological.

In a spinful polar phase with the Hamiltonian (\ref{eq:HamiltonianPolar}) the nodal ring in bulk gives rise to two surface flat bands with opposite chiralities for two directions of spin. The tiny spin-orbit interaction leads to a small splitting of the Andreev-Majorana modes.

\subsection{Flat band on surface of model graphite}

In the multilayered graphene, when the number of the graphene layers tends to infinity, and
if some small matrix elements are neglected, the formed $3+1$ system has line of zeroes, which also obeys the
invariant similar to that in Eq.(\ref{InvariantForLine}). This nodal line has a shape of a spiral
\cite{HeikkilaVolovik2011,HeikkilaKopninVolovik2011}, see Fig. \ref{fig:spiral}.

Let us again consider the momentum ${\bf p}_\perp$ as a parameter of the $1+1$ system, then for
$|{\bf p}_\perp|\neq t$, where $t$ is the dominating hopping element, the system represents the fully gapped system -- the $1+1$ insulators.
This insulator can be described by the same invariant as in Eq.(\ref{InvariantForLine}) with
 the contour of integration chosen parallel to $p_z$, i.e. along the 1D Brillouin zone at fixed
 ${\bf p}_\perp$. Due to periodic boundary conditions, the points $p_z=\pm \pi/a$, where $a$ is the distance between
 the layers, are equivalent and the contour of integrations forms the closed loop.
As a result one obtains the integer valued invariant
\begin{equation}
N_K({\bf p}_\perp)= {1\over 4\pi i} ~{\rm tr} ~\int_{-\pi/a}^{+\pi/a} dp_z ~ \tau_2 H^{-1}\nabla_{p_z} H\,.
\label{InvariantForLine3}
\end{equation}
For  $|{\bf p}_\perp|<t$ the $1+1$ insulator is topological, since $N({|\bf p}_\perp|<t)=1$. This gives rise to the
surface flat band. Since there are no Cooper pair correlations, the fermionic bound states within the flat band are not the Majorana modes.

\section{AM modes on vortices in chiral $2+1$ superfluids}
\label{VortexChiral}

The low-energy fermions bound to the vortex core play the main role in the
thermodynamics and dynamics of the vortex state in superconductors and
Fermi-superfluids. The spectrum of the low-energy bound states in the core of the
axisymmetric vortex with winding number $\nu=\pm 1$ was obtained by Caroli, de Gennes and Matricon  for the isotropic model of
$s$-wave superconductor in the weak coupling limit,
$\Delta \ll \mu$:
\cite{Caroli1964}:
\begin{equation}
E_n(p_z)=-\nu\omega_0(p_z)\left(n+\frac{1}{2}\right)~,
 \label{Caroli}
\end{equation}
This spectrum is two-fold degenerate due to spin degrees of freedom. The
integral  number $n$ is the quantum number related to the  angular momentum of the
bound state fermions. The minigap -- the level spacing $\omega_0(p_z)$ --  corresponds to the
angular velocity of the fermionic quasiparticle orbiting about the vortex axis. The direction of
rotation is determined by the sign of the winding number $\nu$ of the vortex.

The level spacing is typically small compared to the energy gap of the quasiparticles
outside the core, $\omega_0\sim \Delta^2/\mu \ll\Delta$. So, in many physical cases the
discreteness of $n$ can be neglected. In such cases the spectrum crosses zero
energy as a function of continuous angular momentum $L_z$, and one may consider this as spectrum of quasi zero modes. The fermions in this 1D "Fermi liquid" are chiral: the
positive energy fermions have a definite sign of the angular momentum $L_z$.
The number of the branches crossing zero energy as function of continuous $L_z$ obeys the index theorem
\cite{Volovik2003}.

Here we are interested in the fine structure of the spectrum, when its discrete nature is important.
This takes place for example in the ultracold fermionic gases near the Feshbach resonance, when 
$\Delta$ is not small.

 \begin{figure}
 \begin{center}
  \includegraphics[width=1.0\linewidth]{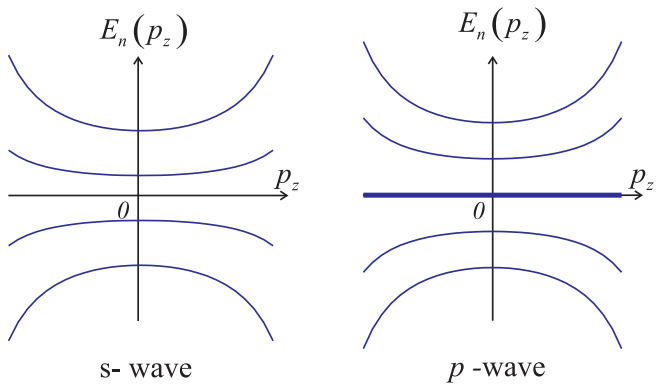}
\end{center}
  \caption{\label{CaroliFigure}
  {\it left} : Schematic illustration of spectrum of  the fermionic bound states in the core
 of $\nu=1$ vortex in $s$-wave superconductor. In the weak coupling limit the lowest branches
are equidistant: $E_n(p_z)=-\omega_0(p_z)\left(n+\frac{1}{2}\right)$. There are no zero energy states.
The spectrum is doubly degenerate over spin.   {\it right} : The spectrum of bound states in the most symmetric vortices in the $p$-wave superfluids: the ciral Weyl superfluid $^3$He-A and the time reversal invariant planar phase.  The spectrum is $E_n(p_z)=-\omega_0(p_z)n$. The branch with $n=0$ forms the flat band of Andreev-Majorana modes.
 }
\end{figure}

Let us first consider the $2+1$ space-time and start with the weak coupling limit.
The Majorana nature of the Bogoliubov particles requires that the spectrum must be symmetric with respect to zero energy,
i.e. for each level with energy $E$ there must be the level with the energy $-E$. For fermions on vortices such condition is satisfied for two
classes of systems. In the systems on the first class the spectrum of Andreev bound states $E_n =\omega_0 (n+1/2)$.
Vortices in $s$-wave superconductors belong to this class. Vortices of the second class  have $E_n =\omega_0 n$. They
contain the Andreev-Majorana mode with exactly zero energy level at $n=0$.
In  the $2+1$ system this mode is not propagating and is self-conjugated. That is why it is called the Majorana
mode instead of the Majorana particle (see Ref. \cite{Wilczek2014}).

Let us for simplicity consider the spinless (or fully spin
polarized) chiral  $p_x+ip_y$ superfluid in $2+1$ space-time,
which is described by Eq.(\ref{2DSpinPolarized}). As it was shown
in Ref. \cite{Volovik1999} the vortices with winding number
$\nu=1$ or $\nu=-1$ belong to the second class:
\begin{equation}
E_n =-\nu\omega_0 n \,,
 \label{NonCaroli}
\end{equation}
and thus contain a single Majorana mode at $n=0$. This mode is
robust to perturbations, since it is self-conjugated and thus must
obey the condition $E=-E$, see also \cite{ReadGreen2000}.

For the spinful fermions in Eq.(\ref{3He-A_film}) there are two
AM modes corresponding to the two spin projections. The even
number of Majorana modes is not robust to
perturbations. For example, the spin-orbit interaction splits
 two modes with $E_1=-E_2$. The splitting is absent, if there is some
discrete symmetry between the AM modes, such as the mirror symmetry in Ref. \cite{SatoYamakageMizushima2014}.

In the spinful
$p_x+ip_y$ there is a topological object, which carries a single
Majorana mode. It is the half-quantum vortex \cite{SalomaaVolovik1987}. In a simple model,
the half-quantum vortex is the vortex with $\nu_\uparrow=1$ in one
spin component, while the other spin component has zero vorticity
$\nu_\downarrow=0$. As  a result such vortex contains single
Majorana mode, which is robust to perturbations.

However, the perturbations should not be too large. In the limit when $\mu$ is negative and large,
the BCS  is transformed to the BEC of molecules, where the Majorana mode is absent. The Majorana mode disappears, when the chemical potential  $\mu$ crosses zero. At $\mu=0$ there is a topological quantum phase transition (TQPT), at which the topological invariant in Eq.(\ref{2+1invariant}) changes from $N=1$ to $N=0$.
The topological transition cannot occur adiabatically, and in the intermediate state with $\mu=0$, the spectrum in bulk becomes gapless. At $\mu=0$ the Majorana mode
merges with the continuous spectrum of bulk quasiparticles and disappears at $\mu<0$.
This demonstrates the topological origin of the AM mode, which exists inside the vortex only if the vacuum in bulk is topologically nontrivial.

\section{AM flat band in a vortex in Weyl superfluids}
\label{VortexFlatBandSec}

One can easily extend the consideration in Sec. \ref{VortexChiral} to the $3+1$ case in the weak coupling limit.
The levels at $p_z\neq 0$ remain equidistant according to the Caroli-de Gennes-Matricon solution and they must be
 symmetric with respect to $E=0$. This dictates the following modification of Eq.(\ref{Caroli}) for the most symmetric
 vortices in $^3$He-A and in the planar phase:
\begin{equation}
E_n(p_z)=-\nu\omega_0(p_z)n \,.
 \label{CaroliExtension}
\end{equation}
This equation suggests the flat band in the vortex core for $n=0$, see Fig. \ref{CaroliFigure} {\it right}.
Now we show how such flat band emerges purely from the topological considerations, which do not use the weak coupling approximation.

Topology of bound states on vortices in  $3+1$ chiral superfluids
can be obtained by dimensional extension of the topology in the
$2+1$ case. The AM mode in the point vortex of fully gapped $2+1$
chiral superfluid transforms to the flat band of AM modes inside
the vortex line in $3+1$ chiral superfluids with Weyl points in
bulk. Let us consider again the $p_x+ip_y$ state in
Eq.(\ref{3He-A_bulk}), and choose for a moment the direction of
the vortex line along the axis $z$. In this case $p_z$ is the
quantum number of the bound states in the vortex core. For each
$p_z$ in the range $-p_F< p_z< p_F$ the Green's function
(\ref{3He-A_bulk}) describes the $2+1$   chiral superfluid with
topological invariant $N(|p_z|<p_F)=1$ in Eq.(\ref{N2+1}), and
this superfluid contains the point vortex. The point vortex in the
$2+1$ topologically nontrivial chiral superfluid contains the AM
mode with zero energy. The continuum of the Andreev-Majorana
modes  in the range $-p_F< p_z< p_F$ forms the flat band.

\begin{figure}[h]
\centering
\includegraphics[width=8cm]{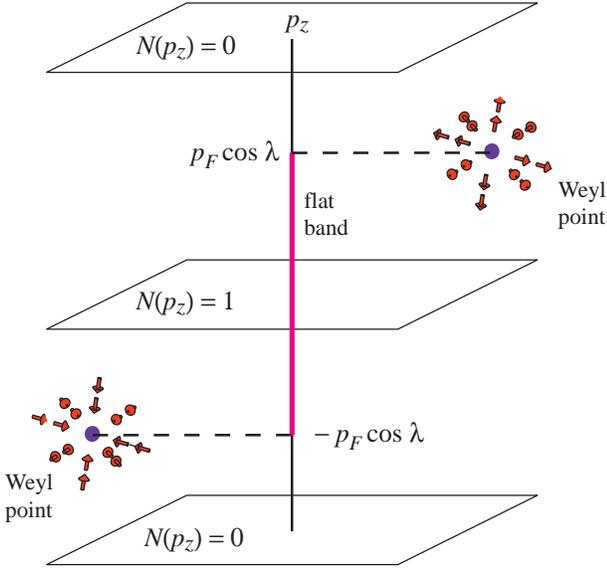}
\caption{
  Projections of Weyl points on the direction of the vortex axis (the $z$-axis) determine the boundaries of the flat band
  in the vortex core. Weyl point in 3D systems represents the hedgehog (Berry phase monopole) in momentum space
\cite{Volovik2003}. For each plane $p_z={\rm const}$ one has the
effective 2D system with the fully gapped energy spectrum
$E_{p_z}(p_x,p_y)$, except for the planes with $p_{z\pm}=\pm p_F
\cos\lambda$, where the energy $E_{p_z}(p_x,p_y)$ has a node due
to the presence of the hedgehogs in these planes. Topological
invariant $N(p_z)$ in \eqref{N2+1} is non-zero for $|p_z| < p_F
|\cos\lambda|$, which means that for any value of the parameter
$p_z$  in this interval the system behaves as a 2D topological
insulator or  2D fully gapped topological superfluid. Point vortex
in such 2D superfluids has fermionic state with exactly zero
energy. For the vortex line in the original 3D system with Fermi
points this corresponds to the dispersionless spectrum of fermion
zero modes in the whole interval $|p_z| < p_F |\cos\lambda|$
(thick line). } \label{fig:vortex}
\end{figure}

\begin{figure*}
\includegraphics[width=0.5\textwidth]{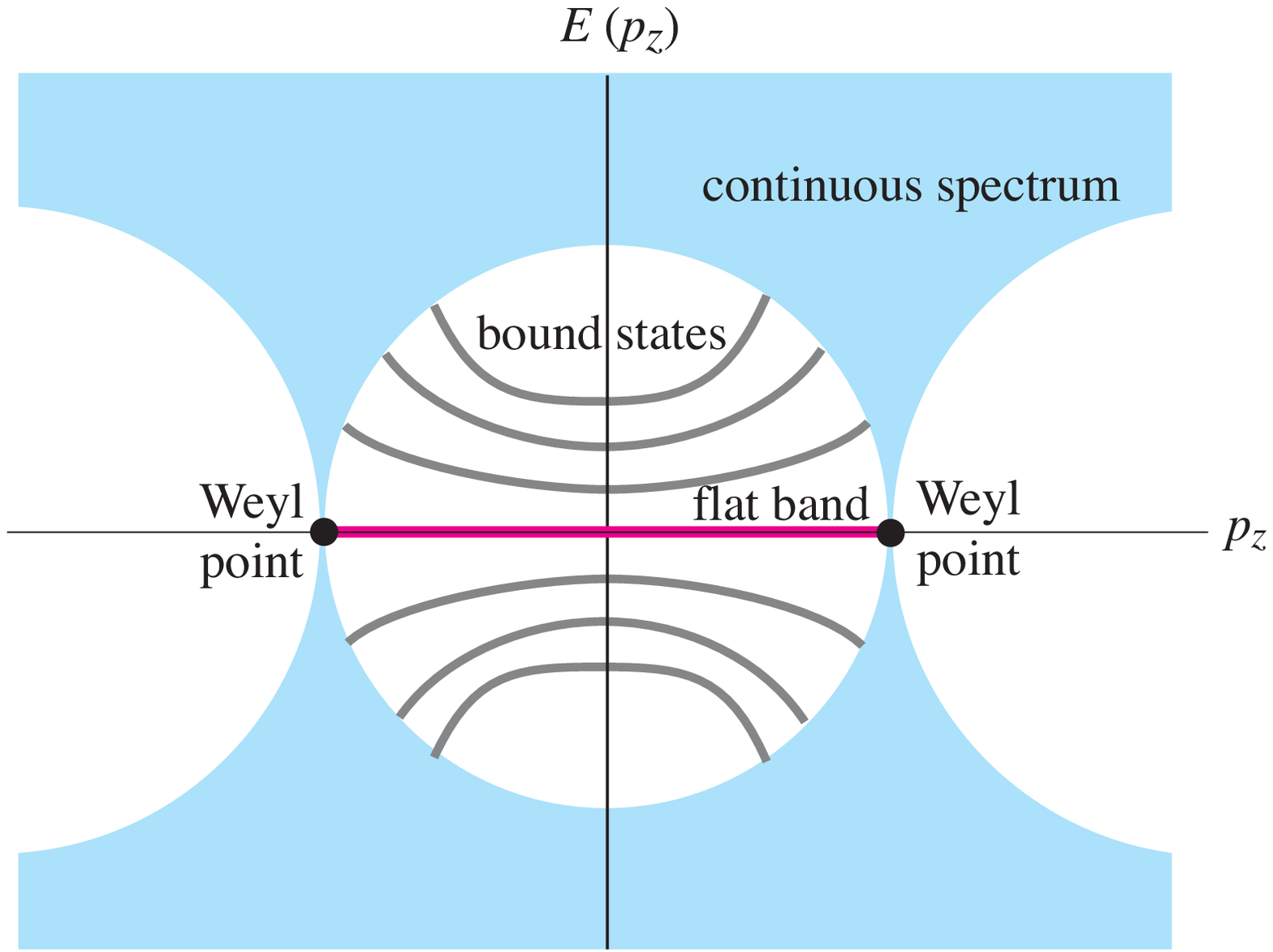}
\caption{Schematic illustration of the spectrum of bound states $E(p_z)$ in the vortex core of Weyl superfluid. The branches of bound states terminate at points where their spectrum merges with the continuous spectrum in the bulk.
The flat band terminates at points where the spectrum has zeroes in the bulk, i.e. when it merges with Weyl points. It is the ${\bf p}$-space analog of a Dirac string terminating on a monopole, another analog is given by  the Fermi arc in Fig. 1 {\it bottom right} .}
\label{VortexFlatBand}
\end{figure*}

This is demonstrated in Fig. (\ref{fig:vortex}), in which the vortex axis is rotated by angle $\lambda$ with respect to the direction to the Wey points. In this case the invariant \eqref{N2+1}
\begin{eqnarray}
 N(p_z)=1~~,~~|p_z| < p_F|\cos \lambda|
 \,,
\label{TopInsulator}
\\
N(p_z)=0~~,~~|p_z| > p_F|\cos \lambda|
 \,.
\label{Non-TopInsulator}
\end{eqnarray}

Such flat band of AM modes has been predicted by Kopnin and Salomaa in Ref. \onlinecite{KopninSalomaa1991} for the $\nu=1$ vortex in $^3$He-A. This flat band is doubly degenerate over spin and thus it may split, for example, due to spin-orbit interaction (the non-degenerate flat band of Andreev-Majorana fermions takes place in the core of half-quantum vortex).  In superfluid $^3$He the spin-orbit interaction is very small and can be neglected. However, there can be another source of splitting:  the symmetry of the vortex core can be spontaneously broken (see \cite{SalomaaVolovik1987}).

The same doubly degenerate flat band should exist in the $\nu=1$ vortex in the $3+1$ planar phase,
where  the Green's function is
\begin{equation}
  G^{-1}= i p_0 +  \tau_3\left(\frac{p^2}{2m} -\mu\right) + \tau_1 (\sigma_x p_x  + \sigma_yp_y)\,.
 \label{planar_state3D}
\end{equation}
Here now $p^2=p_x^2+p_y^2+p_z^2$. For the $3+1$ planar phase, the topological invariant $N_K$
in (\ref{N2+1prime}) is extended to:
 \begin{equation}
 \begin{split}
 &  N_K(p_z)=
 \\
 & =\frac{1}{4\pi^2}~ {\bf tr}\left[ K \int    dp_xdp_ydp_0
 ~G\partial_{p_x} G^{-1} G\partial_{p_y} G^{-1}G\partial_{p_0}
 G^{-1}\right],
 \end{split}
 \label{NK2+1}
 \end{equation}
 giving $N_K(|p_z|<p_F \cos\lambda)=2$.
 
Both flat bands, in the A-phase and in the planar phase, appear only for $\mu>0$, when $N_K(p_z=0)=2$.
For $\mu <0$ the superfluids are  topologically trivial, $N_K(p_z=0)=0$, and the flat band does not exist.

\section{AMBS in $^3$ He-B vortex}
\label{Vortex3HeB}

\subsection{From planar phase to B-phase}
\label{PlanarBphaseVortex}

\begin{figure}
 \begin{center}
 \includegraphics[width=1.0\linewidth]{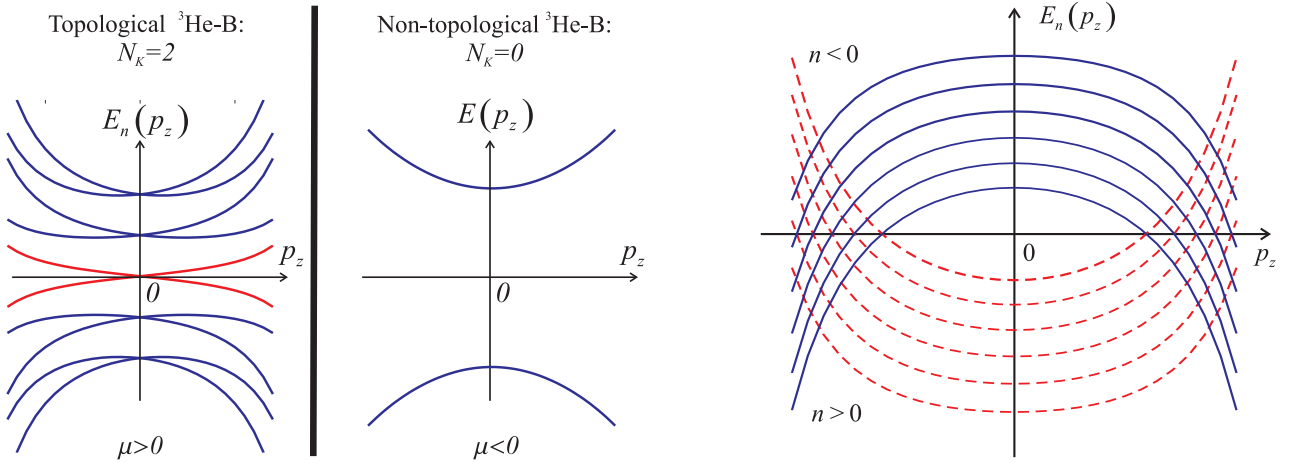}
\end{center}
  \caption{\label{o-Vortex}
({\it a}) Schematic illustration of spectrum of  the fermionic bound states in the core
 of the of the most symmetric vortex ($o$-vortex) in $^3$He-B. Two AM states
with zero energy exist at $p_z=0$.  ({\it b}): The same vortex but in
the topologically trivial state of the liquid, $N_K=0$, does not
have fermion zero modes. The spectrum of bound states is fully
gapped.  Fermion zero modes disappear at the topological quantum
phase transition, which occurs in bulk liquid at $\mu=0$. Similar
situation may take place for strings
 in color superconductors in quark matter \cite{Nishida2010}.
 }
\end{figure}

Dimensional extension of the $2+1$ planar phase allows to
understand the topological properties of the vortex spectrum in
$^3$He-B. The Hamiltonian (\ref{eq:HamiltonianHeB}) for fermions
in the bulk $^3$He-B  represents at $p_z=0$ the $2+1$ planar
phase. That is why at $p_z=0$ the $\nu=1$ vortex in $^3$He-B
contains two AM bound states with zero energy, if the tiny
spin-orbit interaction is neglected and the core symmetry is not
spontaneously broken.  For $p_z\neq 0$ the zero energy modes are
not supported by topology. So the two branches of AM modes split, and 
one may expect the  behavior of the
spectrum of AM bound states in the most symmetric vortex as
illustrated in Fig. \ref{o-Vortex}a,b.

For $^3$He-B, which lives in the range of parameters where
$N^K\neq 0$ in Fig. \ref{o-Vortex}a, the gapless fermions
in the core of the most symmetric vortex (the so-called $o$-vortex
\cite{SalomaaVolovik1987}) have been
  found in Ref. \cite{MisirpashaevVolovik1995}. On the other hand, in the BEC limit, when $\mu$ is negative and the
  Bose condensate of molecules takes place, there are no gapless fermions, see Fig. \ref{o-Vortex}b. Thus in the BCS-BEC crossover  region the
  spectrum of fermions localized on vortices must be reconstructed.  The topological reconstruction of the fermionic
  spectrum in the vortex core cannot occur adiabatically. It should occur only during the topological quantum phase
  transition in bulk, when the bulk gapless state is crossed. Such topological transition occurs at $\mu=0$, see Fig.  \ref{QPT}.
At $\mu<0$  the topological charge $N^K$ nullifies and simultaneously the gap in the spectrum of core fermions arises,
 see Fig. \ref{o-Vortex}b.

This again demonstrates that the existence of fermion zero modes
is closely related to the topological properties of the vacuum
state. The reconstruction of the spectrum of fermion zero modes at
the TQPT in bulk can be also seen
for vortices in relativistic superconductors  \cite{Nishida2010}.

\subsection{AMBS on B-phase vortices with broken symmetry}
\label{Bbroken}

 \begin{figure}
 \begin{center}
 \includegraphics[width=0.8\linewidth]{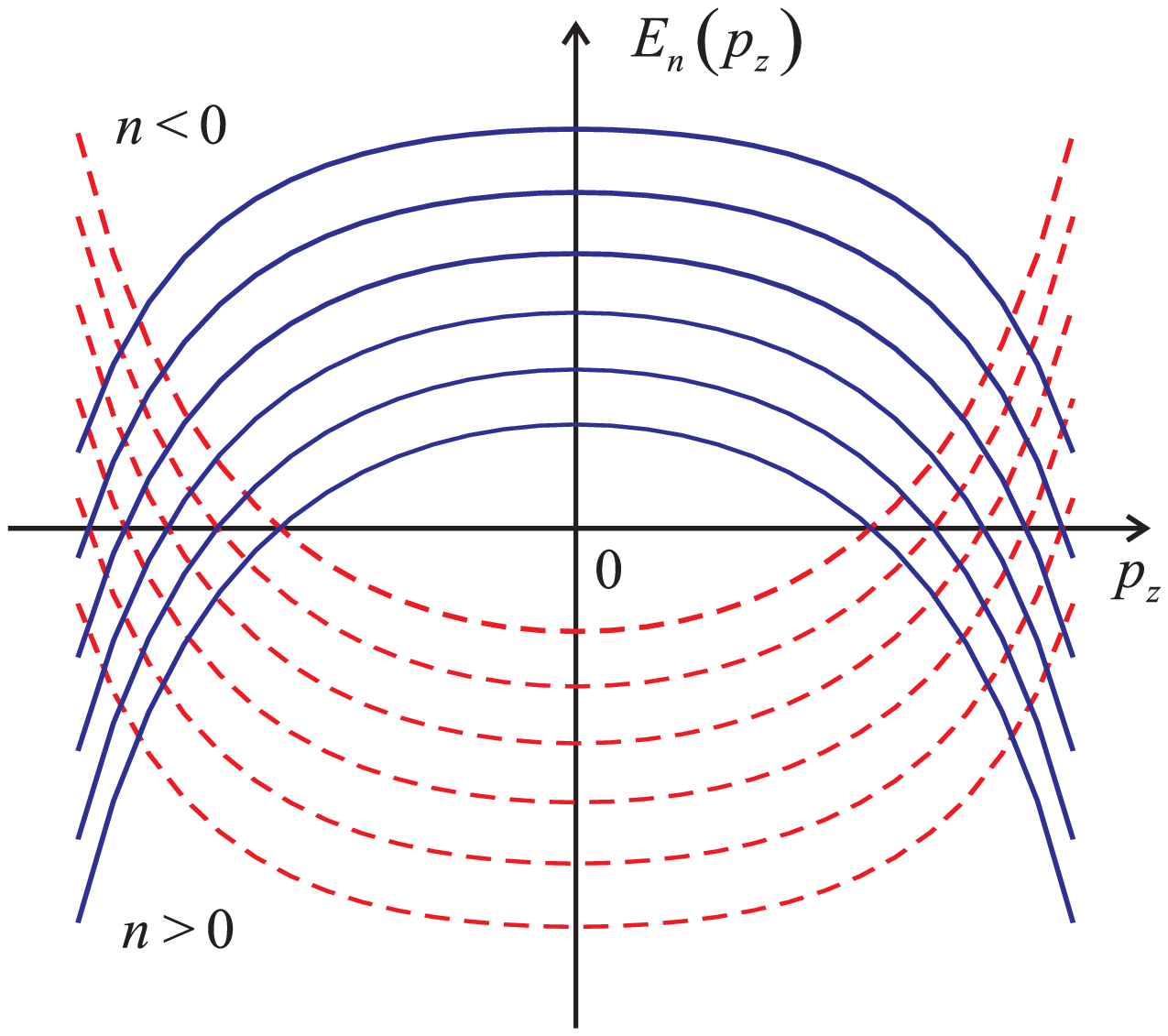}
\end{center}
  \caption{\label{v-Vortex}
  Spectrum of AMBS in the axisymmetric $v$-vortex with
spontaneously broken discrete symmetry in $^3$He-B. The AM states
with zero energy at $p_z=0$, which were present in the most
symmetric $o$-vortex in Fig. \ref{o-Vortex}, do not
exist any more. They split due to matrix element between the spin
components, which appears due to symmetry breaking and move far
away. There are many non-topological branches of spectrum, which
cross zero energy as function of $p_z$ and form
 the one-dimensional Fermi surfaces. The number of such branches $\sim \sqrt{\mu/mc^2}$.
 }
\end{figure}

The spectrum in Fig. \ref{o-Vortex} {\it left} is valid only for
the vortex state, which respects all the possible symmetries of
the vortex core. These symmetries are the space parity $P$ and the
discrete symmetry $TU_2$. The latter is the symmetry under the
time reversal $T$ when it is accompanied by the $\pi$-rotation
$U_2$ about the axis perpendicular to the vortex axis. In the
cores of the experimentally observed vortices in $^3$He-B both
discrete symmetries are spontaneously broken, while the combined
symmetry $PTU_2$ is preserved  \cite{SalomaaVolovik1987}. Such
vortex is called the $v$-vortex. The broken parity in the
$v$-vortex leads to mixing between the two spin components in the
core, as a result the two AM modes at $p_z=0$ split. This leads to
the following spectrum in Fig. \ref{v-Vortex} \cite{Silaev2009}.

In the weak coupling regime, $mc^2 \ll \mu$, there appear the
large number (on the order of $\sqrt{\mu/mc^2}$) of branches,
which cross zero energy. Each crossing point corresponds to the
one-dimensional Fermi-surface. This demonstrates that the topology
in bulk determines the spectrum of the fermion zero modes on the
B-phase vortices only if the symmetry of the vortex core is not
violated. This is the consequence of the mod 2 rule for Majorana
modes: the topological zero energy state survives the symmetry
breaking only in case of the odd number of Majorana modes. So, for
the realistic vortices the AM mode can exist only in half-quantum
vortices.  For the other vortices, such as in $^3$He-B, the large
number of the energy levels is nvolved. That is why for the
consideration it is more appropriate to use the quasiclassical
approximation. The latter leads to the other types of topological
invariants describing the fermion zero modes on vortices, see e.g.
Refs. \onlinecite{Volovik1991b,Volovik2003}.

\section{Conclusion}
\label{sec:conclusion}

We considered the Andreev-Majorana bound states with zero energy on surfaces, interfaces and vortices in
different phases of the $p$-wave superfluids: $^3$He-A, $^3$He-B, planar and polar phases.

These states are determined by topology in bulk, and they disappear at the quantum phase transition from the topological
to non-topological state of the superfluid (see example in Fig. \ref{o-Vortex}). This topology demonstrates the interplay of dimensions.
In particular, the 0D Weyl point (the Berry phase monopole in momentum space) gives rise to the 1D  Fermi arc on the surface
(Sec. \ref{FermiArcBoundary}). The 1D nodal line in bulk produces the dispersional 2D band of AM modes on the surface (Sec. \ref{PolarFlatBand}).

The interplay of dimensions also connects the AM states in superfluids in different dimensions.
For example, the property of the spectrum of bound states in the 3D  $^3$He-B is connected to the property of the spectrum in
the 2D planar phase (see Sec. \ref{AMBSgapped} for edge states and
Sec. \ref{PlanarBphaseVortex} for bound states on vortices). The 0D AM mode on a point vortex in 2D chiral superfluid
(Sec. \ref{VortexChiral}) gives rise to the 1D flat band of AM modes on a vortex in the 3D chiral superfluid
(Sec. \ref{VortexFlatBandSec}).

The most robust zero energy edge states take place on the boundary of $^3$He-A, or in general on boundaries and  interfaces of
 chiral superfluids with the topological invariant $N$ in Eq.(\ref{2+1invariant}). In the other phases,
 the existence zero energy edge states is supported by symmetry, i.e. by the symmetry protected topological
 invariants $N_K$ in Eqs. (\ref{N2+1prime}) and (\ref{InvariantForLine}).
When the symmetry is violated in bulk or on the boundary/interface, the AM bound states acquire gap.

Concerning the AM states on vortices, only the states on half
quantum vortices are fully robust to perturbations. In the singly
quantized vortices the fate of zero energy states depend on
symmetry and its possible violation in bulk or spontaneous
breaking inside the vortex core. This the consequence of the $Z_2$
classification of the AM modes on vortices. On the other hand, the
spontaneously broken symmetry inside the vortex core may give rise
to many non-topological branches of AMBS, which cross zero energy
as function of $p_z$. This is demonstrated in Sec. \ref{Bbroken}.

Let us also mention the application to the relativistic theories.
The fermion zero modes obtained in the Dirac systems, such as the
modes localized on strings in Ref. \cite{JackiwRossi1981}, are not
properly supported by topology. The reason for that is that the
Dirac vacuum is marginal, and the topological invariants depend on
the regularization of the Green's function in the ultraviolet
\cite{Volovik2010a}. For example, in  Fig.\ref{QPT} the Dirac
vacuum is on the border between the trivial vacuum with $N_K=0$
and the topological vacuum with $N_K=2$. That is why the existence
of the modes with exactly zero energy depends on the behavior of
the Green's function at infinity.

\section{Acknowledgements} \label{Sec:Acknowledgements}
MS and GEV acknowledge financial support  by the Academy of
Finland through its LTQ CoE grant (project $\#$250280).

\end{document}